# The Social and Psychological Impact of Cyber-Attacks

Maria Bada
Cybercrime Centre, Computer Laboratory, University of Cambridge, UK
maria.bada@cl.cam.ac.uk

Jason R. C. Nurse
School of Computing, University of Kent, UK
j.r.c.nurse@kent.ac.uk

**Abstract:**
Cyber-attacks have become as commonplace as the Internet itself. Each year, industry reports, media outlets and academic articles highlight this increased prevalence, spanning both the amount and variety of attacks and cybercrimes. In this article, we seek to further advance discussions on cyber threats, cognitive vulnerabilities and cyberpsychology through a critical reflection on the social and psychological aspects related to cyber-attacks. In particular, we are interested in understanding how members of the public perceive and engage with risk and how they are impacted during and after a cyber-attack has occurred. This research focuses on key cognitive issues relevant to comprehending public reactions to malicious cyber events including risk perception, protection motivation, culture, and attacker characteristics (e.g., attacker identity, target identity and scale of attack). To consider the applicability of our findings, we investigate two significant cyber-attacks over the last few years, namely the WannaCry attack of 2017 and the Lloyds Banking Group attack in the same year.

## 1. Introduction

The impact of cyberspace on society is undeniable. It has provided a platform for instantaneous communication, commerce and interaction between individuals and organizations across the globe. As cyberspace has grown in prominence however, unfortunately so too has the number and variety of cyber-attacks (Verizon, 2018). Cyber-attacks are defined here as events which aim to compromise the integrity, confidentiality or availability of a system (technical or socio-technical). These attacks range from hacking and denial-of-services (DoS), to ransomware and spyware infections, and can affect everyone from the public to the critical national infrastructure of a country (Nurse, 2018). In this article, we will examine this topic of cyber-attacks from more of a social and behavioral science perspective, with the aim of exploring the social and psychological factors and impacts associated with these attacks.

Research has shown that members of the public are more likely to respond to the effects of a cyber-attack rather than the attack itself (Minei & Matusitz, 2011; Gandhi, Sharma, Mahoney, Sousan, Zhu & Laplante, 2011). One example of this is a cyber-attack where malware infects a national power station causing the hundreds of thousands of citizens to be without power. Here, the attack, i.e., the malware infestation may not worry individuals (the public), they will be much more considered about the effect, i.e., being without power, thereby having no heating, ability to prepare food and so on. There are two key areas of impact that we aim to consider and provide an overview of the current research and thinking; these are the social and psychological (emotional and behavioral) impacts. The social impact of a cyber-attack refers to aspects such as the social disruption caused to people's daily lives, and widespread issues such as anxiety or loss of confidence in cyber or technology. Psychological impact can be informed by social impact, and can include more personal aspects such as an individual's anxiety, worry, anger, outrage, depression and so on.

To inform our research, the chapter begins by first reflecting on some of the key issues relevant to understanding public reactions to malicious cyber-events. We examine topics such as: risk perception (Nurse, Creese, Goldsmith & Lamberts, 2011; Rogers, Amlôt, Rubin, Wessely & Krieger, 2007); locus of control (Ajzen, 2002); culture of fear (Stekel, 1930); the online disinhibition effect (Suler, 2004); and protection motivation (Rogers, 1975; Maddux & Rogers, 1983; Blythe & Camp, 2012); amongst others. Also, within scope of our assessment are the range of potential factors which can influence the public's level of perceived risk, such as the perpetrator's identity and the scale of the cyber-attack.

Beliefs form an important component of our investigation. This is because a user's reaction to security generally and motivation to apply security mechanisms – if given the chance – depends on their beliefs about: the perceived severity of an event; the susceptibility to the threat; the perceived self-efficacy; and the cost and efficacy of preventative or mitigating behaviors (Blythe & Camp, 2012). These factors make it difficult to motivate protective cybersecurity practices (behaviors) as well as to predict public social and psychological responses to a cyber-attack. This can be any form of attack ranging from existing threats to new concerns such as the cyber risks with Internet of Things or Artificial Intelligence (Nurse, Creese & De Roure, 2017).

Another element of relevance is the general culture of fear related to crime and cyber-events. Fear of crime can prompt people to change their behavior. At the level of the individual, people generally respond to the fear of crime by adopting protective or avoidance behaviors (Reid, Roberts & Hilliard, 1998). Phobophobia – the psychological fear of fears (Furedi, 2002) – can lead to stress, intense anxiety, and unrealistic and persistent public fear of crime and danger, regardless of the actual presence of such fear factors. This phenomenon may also relate to the crime complex (Hale, 1996) and therefore cyber-attacks and cybercrime.

Having discussed risk and attack perceptions, protection motivation theories and theories regarding public reactions to attacks and online crimes, we then critically reflect on two real-world cyber-attack scenarios from 2017: the global WannaCry attack and the cyber-attack on the Lloyds Banking Group. This seeks to understand the social and psychological impacts resulting from these attacks on an individual basis as well as to the wider society. These are all important topics for discussion and analysis as we aim to advance research into cyber threats from a cyberpsychology perspective.

## 2. Factors influencing perceptions of risk and reactions to risk

A user's motivation to react to perceived risk and apply security measures depends on their beliefs about: the perceived severity of an event; the susceptibility to the threat; the perceived self-efficacy; and the cost and efficacy of preventative or mitigating behaviors (Blythe & Camp, 2012). Also, the general culture of fear related to crime and cyber-events (Furedi, 2002) can prompt people to change their behavior. In this section, the factors influencing perceptions of risk related to cyber events and reactions to risk are discussed. Moreover, different theories and theoretical models are presented as related to conduct online.

### 2.1 Perception of risk
Research in public perception of risk (Slovic, 1998, 2000; Sjöberg, 2000; Dickert, Västfjäll, Mauro & Slovic, 2015) demonstrates that there are potential factors which can influence the public levels of perceived risk such as whether or not exposure to the risk is perceived to be: (i) voluntary (accepting increased risk through risky online activities or involuntary (as opposed to knowingly accepted); (ii) familiar (due to the frequency of appearance in the media) or unfamiliar (i.e. lack of understanding of the causes and consequences); (iii) controllable (due to safeguards that can be put in place) or uncontrollable (as opposed to feeling in control); (iv) fair (i.e. random) or unfair (i.e. targeted); and (v) whether or not the risk causes 'dread'.

Also, it is suggested that attitude, risk sensitivity, and specific fear can be used as explanatory variables for risk perception (Sjöberg, 2000). Nurse et al. (2011) further evidence this in considering that persons use four main dimensions in judging online risks, namely ability to control or avoid the risk, dread of consequences, unfamiliarity of risks and immediacy of consequences/impact. Members of the public acknowledge the threat of cyber-attacks, but the steps that they take to address this threat vary. People react to risk in different ways based on dual information processing. Some react based on logic, analyzing risk, and others might react instinctual based on feelings about the risk (Dickert et al., 2015). Emotions can also serve as a spotlight for directing our attention but also motivating individuals to act. For example, people can decide on acting on risks related to technologies based on their feelings toward specific outcomes (Nurse, 2018).

Other authors propose that, both experts and members of the public might confuse facts with their individual interpretation because perceptions of risk are often based on the interpretation of facts, which are fed by individual judgement, values, beliefs and attitudes (Beck, 1999). Blythe & Camp (2012) argued that overall, a user's motivation to apply security mechanisms depends on their beliefs about their susceptibility to exogenous security threats, their potential severity and the cost and efficacy of preventative or mitigating behaviors. This therefore fits with existing and prior literature.

Security decisions and behaviors are executed in a world of risk and uncertainty. Adams (2013) explains the notion of risk compensation by presenting a risk thermostat. He claims that individuals execute a balancing behavior between their propensity to take risks (risk appetite) and perceived danger (risk perception), where risk propensity is determined by perceived rewards, whereas accidents (negative experiences) influence perceived danger.

Trust has been identified as a key issue impacting public perceptions of risk. The level of trust in an organizational body responsible for responding to the risk should be considered during both the policy-making and communication processes (Rogers, Amlôt, Rubin, Wessely & Krieger, 2007). For example, a report by Symantec (2010) showed that nearly 9 in 10 adults are considering cybercrime and over a quarter actually expect to be scammed or defrauded online. Yet despite the universal threat and incidence of cybercrime, only half of adults in the study claim that they would change the way they behave online if they became a victim.

Therefore, policy makers need to understand how people think about and respond to risk and materialized attacks, given that without such insight policies or awareness efforts might be unsuccessful.

**2.2 Self-efficacy**
Self-efficacy is considered not as skills themselves, but as the evaluation of what one can do with skills. It considers a person's belief in themselves and their abilities (Bandura, 1991). This concerns issues such as cyber-attacks because it is important that individuals believe that they stand a chance of protecting themselves and responding successfully to an attack's occurrence.

Ajzen (2002) introduced a new concept about the relationship between self-efficacy and perceived behavioral control. He argued that "*the central concept of perceived behavioural control consists of two factors: self-efficacy (about the ease / difficulty of performing a behavior) and the ability to control (the extent to which performance depends exclusively on the individual)*". When individuals are able to determine or influence what is happening to them or what will happen to them, these individuals are considered to "be under control". Control is a central construction in psychology, and being under control is a worldwide desirable state of being for most people. The same reality holds in the online world, generally and in the face of a spate of emerging cyber-attacks.

**2.3 Protection motivation theory (PMT)**
According to Protection Motivation Theory (PMT) (Rogers, 1975, 1997), environmental and personal factors are combined to pose a potential threat. The threat initiates two cognitive processes: threat appraisal and coping appraisal. The threat appraisal process evaluates the factors associated with the behavior that potentially creates danger, including the intrinsic and extrinsic rewards accompanying the actions, the severity of the danger, and one's vulnerability to the threat. The coping appraisal process evaluates one's ability to cope with, and avert, the threatened danger (self-efficacy and response efficacy), balanced with the costs (or efforts) associated with protective behavior (response cost).

Threat appraisal refers to how susceptible one feels to a threat. For example, how vulnerable is an individual by the possibility of becoming a victim of a cyber-attack such as phishing; naturally, susceptibility to phishing attacks is influenced by a range of other aspects (Iuga, Nurse & Erola, 2016; Williams, Hinds & Joinson, 2018). Coping appraisal evaluates the various factors that are likely to ensure that one engages in a recommended response that is preventive in nature. For instance, not opening emails being sent by an unknown sender, or untrustworthy or suspicious email address.

The theory therefore says that in order for an individual to adopt a safe behavior, they need to believe that there is a severe threat that is likely to occur and that by adopting safe actions, they can effectively reduce the threat. The individual should also be convinced that they are capable of engaging in the behavior and that it would cost only expected amounts in terms of effort expended. Measuring the 'intention' to engage in the recommended preventive activity is the most common indication of protection motivation.

Another key aspect of understanding the public response to malicious cyber-attacks centers upon the fact that members of the public do not appear to perceive such attacks as a threat to themselves and, if they do, they believe that there is very little that they can do to prevent such an attack. Instead, members of the public are more likely to respond to the event (e.g., loss of service), rather than the cyber-attack, itself. We discuss this further in Section 4. Perceptions of what others expect or how others react to a threat are aspects of both types of appraisal. In general, increases in threat severity, threat vulnerability, response efficacy, and self-efficacy facilitated adaptive intentions or behaviors. Conversely, decreases in maladaptive response rewards and adaptive response costs increased adaptive intentions or behaviors.

These factors make it difficult to infer social and psychological responses to a cyber-attack. If we interpret malicious cyber-attacks through the lens of PMT, the current state of the public understanding of the potential impact of such an event would possibly be described in high levels of threat appraisal, low levels of self-efficacy, confusion over response efficacy, and high response costs due to the perceived difficulty of enacting security measures.

## 2.4 Locus of Control

Another theory that can be utilized to describe emotional and behavioral responses to an online incident is that of locus of control (Ajzen, 2002). Locus of control aims to characterize whether people feel they have strong control over their life (internal locus of control) or whether they have to rely on external forces (external locus of control). The locus of control appears to affect learning, motivation and behavior.

Persons with an internal locus of control feel that success or failure is due to their efforts or abilities. Alternately, individuals with an external locus of control are likely to believe that other factors such as luck or the difficulty of the target or the actions of other people are the cause of success or failure. For example, users with external locus of control might not often take protective measures against cyber-attacks due to their beliefs that Internet providers or the government is responsible for ensuring a safer Internet.
Lack of control over a situation that is perceived as threatening or dangerous can give rise to feelings of emotional distress, fear and insecurity. Such strong emotions can on occasion lead to irrational behavior (Sutherland, 2007) or other equally strong reactions. These are all aspects that can influence how members of the public (psychological) and society generally (social) are impacted by an attack.

## 2.5 Extended Parallel Process Model

Similar to PMT, the Extended Parallel Process Model (Witte, 1992) suggests that when the perceived threat is low, independent from the level of perceived response efficacy, there may be no further processing of the message. Thus, there is no reaction to the invocation of fear because the threat is not subject to further processing (Witte, 1991). Also, the Model proposes that as the perceived threat grows while the counter-effect is high, the acceptance of a preventive advice will also increase. In such cases, individuals can realize that they are at risk of a serious threat and are motivated to protect themselves; consider that they can prevent the risk (high efficacy); and may deliberately and cognitively take action to address the risk. Fear invocations with high levels of threat and high levels of efficacy produce acceptance of a suggested behavior (Kleinot & Rogers, 1988; Maddux & Rogers, 1983, Rogers & Mewborn, 1976, Witte, 1992).

The cognitive processes that take place during risk control procedures trigger adjustment actions such as attitudes, intentions or behavioral changes that control the risk. However, as the perceived threat grows while the perceived efficacy is low, individuals will do the opposite of what is proposed. There is the argument that in order to control the unbearable fear of a state of low perceived efficiency, a person will either consciously or unconsciously, tend to deny the threat, or react against the suggested preventive behavior, and perform even more risky behavior to reduce fear or anxiety. This can relate to both the offline and the online environments. Overall, when the severity of the threat is high, combined with low efficacy, then people may tend to reject the suggested actions or are led to boomerang reactions (Kleinot & Rogers, 1982; Rippetoe & Rogers, 1987; Rogers & Mewborn, 1976; Witte, 1992).

These points further evidence research around the effectiveness of cybersecurity awareness campaigns which often leads to failure of behavior change (Bada, Sasse & Nurse, 2015). In addition, when the perceived effectiveness is moderate, the critical point may not occur directly, but at a moderate level of the threat. For example, when perceived efficacy is moderate, people may initially think they can prevent a cyber-attack. However, as the threat grows in intensity and relativity, individuals may begin to give up any hope of averting the threat or adequately addressing any subsequent impact.

**2.6 Culture of Fear**
In the psychology of fear literature, Stekel (1930) makes the point that fear is in part hereditary, a legacy of centuries, that left its traces in our brain. Correspondingly, perceived efficacy consists of the individual's views on the severity of the threat, while perceived vulnerability consists of the person's attitudes about their chances of dealing with it.

The theory of self-efficacy (Bandura, 1977, 1986) presupposes that the perceived ineffectiveness of dealing with possible events is one that creates both fear expectations and avoidance behavior. Individuals who judge themselves to be effective in managing potential threats, may feel neither fear nor avoid threats. On the contrary, if people judge themselves as ineffective in exercising control over potential threats, they react with stress and do not want to have any contact with them, therefore avoiding them. For example, in the occasion of a cyber-related incident such as a phishing scam, individuals might judge themselves as not having the necessary skills or knowledge in order to avoid such an incident, therefore avoiding to act or take any protective actions. If this is to occur on a large scale, it could have a notable social impact.

Fear expectations and avoidance behaviors are factors that can influence perceived inefficiency of managing situations. Acknowledging that human behavior is largely regulated by personal efficacy beliefs, people can exercise their activities at the lowest levels of self-efficacy despite high fear invocation and can take preventive actions without having to wait for the feelings of fear and excitement to arise. Different theoretical approaches explain fear control procedures (Leventhal, 1970) or the way individuals cognitively recognize fear or threat, changing their attitudes, intentions or behaviors to avoid the threat (factors leading to acceptance of the message). Examining the causes of fear reveals some interesting realities. According to Witte (1994) the greater the threat, the greater fear is expressed. Also, the threat is related to the sense of fear and not to efficiency (Witte, 1994).

Witte (1992a) argued that perceived effectiveness only determines the nature of the reaction (control of fear or risk), while the perceived threat determines the intensity of the reaction (how much control of fear or risk is caused). Considering that the perceived threat and the fear-causing appear to be closely related, it is likely that the perceived threat and the emotion of fear cooperate to influence the intensity of a reaction to a call of fear. To place this in the context of online threats, when individuals are afraid of a major cyber-attack and realize that a reaction could effectively prevent that threat, they are motivated to control the risk (protection incentive). This control process could begin with thinking about strategies to tackle that threat and reduce the impact of the corresponding risk on their lives. When risk control procedures dominate, people may arguably react to danger, not to fear.

Considering our case of cyber-attacks, another example would be that victims of fraud and computer misuse who have been victimized before might take measures to avoid becoming a victim again in the future. The vast majority of victims of fraud and computer misuse have only been victimized once, with only a small proportion saying they have suffered two or more times. Statistics support this claim indicating that users can be quick to learn from their mistakes when they become victims of computer crime (Reeve, 2017).

Conversely, when perceived threat is high, but the perceived effectiveness is low, the fear control procedures are initiated. Fear is initially caused, and the threat becomes intense when individuals feel unable to prevent the threat. Thus, they are mobilized to manipulate their fear (defensive manipulation) by adopting reactions, such as denial. When fear control procedures dominate, individuals react to their fear, not to danger. Individuals might feel helpless and victimized while their lack of knowledge about cybercrime will lead them accepting the possibility of being victims or denying that possibility overall. Moreover, victims feel the usual set of emotions when they realize they have been scammed – from helplessness to rage.

According to Garland (2001), when it comes to fear of crime, "*our fears and resentments, but also our common-sense narratives and understandings, become settled cultural facts that are sustained and reproduced by cultural scripts*". The idea of 'cultural scripts' can help to reveal much about emotions such as fear; arguably even in the cyber-attack context. A cultural script communicates rules about feelings, and also ideas about what those feelings mean. People interpret and internalize these rules according to their circumstances and temperament, while always remaining very much influenced by the rules. Consequently, the impact of fear is determined by the situation people find themselves in, but it is also, to some extent, the product of social construction (Altheide, 2002). Fear is determined by the self, and the interaction of the self with others; it is also shaped by a cultural script that instructs people on how to respond to threats to their security.

## 2.7 The Online Disinhibition Effect

Regarding the impact of cyber-attacks on online behavior there are different aspects of cyberspace that need to be considered. One of these aspects surrounds the reality that individuals say and do things in cyberspace that they would not ordinarily say and do in the offline (face-to-face) world. For instance, they may loosen up, feel less restrained, and present themselves more openly. So pervasive is this phenomenon that a term has surfaced for it, namely the online disinhibition effect (Suler, 2004).

This disinhibition is empowered due to several factors. These consider the fact that people can form a different identity online and that they may feel less vulnerable in the way they express themselves or behave, in comparison to how they would act offline. Moreover, people can feel less visible online and therefore might engage in activities that they otherwise would not. This factor affects our discussions on impacts as the consequence of some online actions or activities may not be fully tangible to people in the offline world. This has also been discussed in other areas as it relates to cybercrime (Nurse, 2018).

## 3. Understanding Public Reactions to Malicious Cyber Incidents

### 3.1 Emotional Reactions to Cybercrime

Research indicates that current forms of cyber-attacks can cause psychological impacts (Gandhi, et al., 2011; Dallaway, 2016; Modic & Anderson, 2015). Depending on who the attackers and the victims are, the psychological effects of cyber threats may even rival those of traditional terrorism (Gross, Canetti & Vashdi, 2016). Victims of online attacks and crime can suffer emotional trauma which can lead to depression. There is also some evidence of limited symptoms of Acute Stress Disorder (ASD) in victims of crime in online virtual worlds, such as some anecdotal accounts of intrusive memories, emotional numbing and upset from victims of virtual sexual assault (Lynn, 2007).

As an example, the impact of identity theft on a victim at an emotional level can lead the person becoming distressed and be left feeling violated, betrayed, vulnerable, angry and powerless (Kirwan and Power, 2011). Often, victimization can lead victims to feelings of outrage, anxiety, a preference for security over liberty, and little interest of adopting new technology due to loss of confidence in cyber. The victim can go into stages of grief, suffer from anger or rage. In some cases, victims may even blame themselves and develop a sense of shame; sextortion is a good example of this given how it initially starts (Nurse, 2018).

In other work, a Symantec (2010) study further showed that victims feel that they themselves are partly or wholly to blame; this in itself has consequences for the resulting psychological impact. The number who act varies widely depending on location. For instance, 74% of those in Sweden contact the police, but this is significantly above the overall average of 44%. As a general rule, around half of victims will not contact anyone, although about a quarter might try taking some action themselves, even if it is only avoiding certain websites in the future. Other impacts can be isolation or even depression especially in the event of a financial loss.

According to Symantec, the top 10 emotional reactions to online attacks and cybercrime are the feelings of anger, annoyance and being cheated. Böhme and Moore (2012) found that directly experiencing cybercrime decreases the likelihood of shopping and banking online, while expressing concern about cybercrime has nearly twice as much negative impact on online behavior than directly experiencing it. Modic and Anderson (2015) found that victims of financial fraud consistently reported emotional impact as more severe than financial impact across all fraud types.

Reflecting on the literature, we can see that it also demonstrates how even non-lethal forms of cyber terrorism have a considerable impact on the attitudes of victimized populations (Gross, Canetti and Vashdi, 2016). Under attack, victims react with not only fear, as do victims of crime, but with demands for protection from the government, via surveillance and stronger regulations.

Estonia is frequently invoked in such discussions and a host of authors draw attention to the panic caused amongst the people of Estonia when parts of their cyber infrastructure were inaccessible due to DoS attacks' in 2007 (Gandhi, Sharma, Mahoney, Sousan, Zhu and Laplante, 2011). However, the potential of large-scale malicious cyber-attacks to change the public understanding and perception of cyber events could lead to an activation of the 'dread factors' of risk perception (i.e. catastrophic potential, fatal consequences, and high risks to future generations) which, in turn, can inform a multitude of spontaneous precautionary behaviors. This becomes of particular concern given the several predictions of possible future technology scenarios, and implications for security and privacy online (Williams, Axon, Nurse & Creese, 2016).

**3.2 Learned Helplessness**
Findings show that less than 1 in 10 people (9%) claim that they feel 'very' safe online. Also, only half (51%) of adults asked, would change the way they behave online if they became a victim (Symantec, 2010). This provides an interesting comparison to our earlier discussions. People might accept a situation, even if it feels unpleasant just because they cannot understand it or do not know enough about it. Following this point, one might argue that persons may accept cyber-attacks because of a sense of 'learned helplessness'.

Due to a sense of learned helplessness (for more on the term, see: Seligman, 1975; Hirtz, 1998) and a lack of knowledge about online attacks and ways to resolve an incident, users may simply accept the possibility of being victims. Indirectly, a key question therefore becomes, whether they also accept the reality of impacts and hope that the severity is low. The anonymous nature of cybercrime, can lead to an acceptance that one (e.g., an individual, industry, government) will become a victim of cybercrime at some point. Moreover, the sense of learned helplessness can potentially also result in a low uptake of protective security behaviors.

Users are called upon to make many security-related decisions every day which can cause anxiety. These behaviors include: (a) not opening an email from a sender they do not recognize; (b) not accessing unknown attachments; (c) only downloading and running programs from trustworthy sources; (d) the use of anti-virus software and security software (e.g., firewall); and (e) creating regular backups. Some of these decisions can also cause the user feelings of anxiety due to a lack of knowledge about the possible implications of making incorrect decisions.

Members of the public have often reported a lack of knowledge about a number of key areas within the cybersecurity domain. A few examples are lack of knowledge about how to use security packages, how to secure their technology devices, and the threats online. Even when these individuals are aware of the threats, they may report that they do not understand them (Gross, Canetti and Vashdi, 2017). These low levels of public understanding of cyber-threats and security practices could lead to a lack of public engagement with security issues and a general loss of confidence in cyber and/or technology. This also has been seen in the domain of information privacy in the context of new forms of technology, where some users now consider privacy as 'the boring bit' (Williams, Nurse & Creese, 2017). These issues characterize the broad social impacts.

### 3.3 Cyber-attack related variables
The public response to a cyber-attack is informed by a number of cyber-specific variables such as the attacker identity, the target identity, the scale of the attack as well as the government communication of a cyber-attack and the time of revelation of a malicious event.

Public reactions may differ according to the disclosed identity of a given attacker. The principal categories of an actor are terrorist, hacktivist, and criminal – all of whom might be capable of launching attacks that could qualify as issues of serious public concern (Nurse & Bada, 2018). Criminals are, on average, less likely to publicly reveal their identity (assuming any identity, pseudonym or otherwise) because anonymity facilitates them better. Moreover, the target identity can impact the public response. For example, if a series of fraud incidents impacts individuals at random, it may be expected that it will cause less panic or outrage as compared to a targeted attack towards a national financial, utility or health institution.

Additionally, the scale of an attack will influence its impact. The full extent of an attack might not become apparent immediately, particularly if second and third-order systems fail. Finally, the way that the government will communicate a cyber-attack and the time of revelation of a malicious event will impact the level of public response. This information can influence the direction and dynamics of public response. The ways in which members of the public are likely to find out about a cyber-attack is also an important variable. Different levels of public response can be caused due to loss of service, public announcements from the attacker or from government announcements, as will be seen in the next section of this chapter.

Lawson (2013) draws upon the history of technology and failures of large sociotechnical systems, military history and—most pertinent here—disaster sociology to suggest that 'fear and panic' may not be the defining features of public reactions to future cyber-attacks. As mentioned above, members of the public are more likely to respond to the event (e.g. loss of service), rather than the cyber-attack itself.

**4. Case Studies of Cyber-Attacks**

Thus far in this chapter, we have reflected upon the literature pertinent to understanding the impact of cyber-attacks at the psychological and social levels. We now extend this discussion to real-world cases of such threats that have occurred over recent years. In particular, this work focuses on two cyber-attacks in 2017, namely the WannaCry attack and the cyber-attack on the Lloyds Banking Group. Our aim is to examine these attacks and the impacts that they have had from a social and psychological perspective, while drawing on insights presented earlier in this chapter.

**4.1 WannaCry ransomware attack in 2017**
WannaCry was a computer worm responsible for one of the most devastating cyber-attacks in recent history. In addition to spreading across computer networks using the Windows operating system, WannaCry (also known as WannaCrypt and WanaCrypt0r) encrypted files of the host computer and only allowed access to those files after a bitcoin ransom payment was made. While ransomware itself was not new, WannaCry was particularly successful because it targeted a computer vulnerability that many organizations and individuals did not yet install security patches (i.e., software updates) to address.

Reflecting on the social impact of the attack, WannaCry infected over 200,000 victims in at least 150 countries (Reuters, 2017). These included members of the public, but also healthcare organizations, car manufacturers, telecoms companies, delivery services and the education sector. Due to the nature of the attack, the disruption it caused at the social level was quite significant. Organizations closed (causing people to be sent home), production stopped (resulting in product backlogs), and many businesses were unaware of how best to restore services. Overall, people felt a loss of control (Ajzen, 2002) as the threat was so pervasive and the only option for recovery – assuming no recent backups were made – was to pay the ransom. In total, these disruptions led to an estimated $8 billion in economic costs globally (Barlyn, 2017).

In the UK, critical infrastructure such as the National Health Service (NHS) was also impacted. A total of 48 NHS Trusts were infected in England and 13 in Scotland (BBC, 2017a). This resulted in direct disruption to people's lives in the context of their health. Specifically, there were cancelled operations, certain scans and treatments were postponed, and ambulances were diverted. This is a particularly worry situation given how long patients often have to wait for some treatments. WannaCry also prompted government responses – in the UK a Cobra meeting (i.e., an emergency response government committee meeting) was called, and in the US, the homeland security adviser, was ordered to coordinate the government's response and help to organize the search for the responsible parties (Sanger, Chan, and Scott, 2017). This level of response from governments further emphasized the impact of the threat in the eyes of the public.

An interesting point to note here is the government's comments on the attack. In particular, there was an emphasis on the fact that the attack had not targeted the NHS and that it was international. According to The Independent, the UK's Home Secretary stated: "*If you look at who's been impacted by this virus, it's a huge variety across different industries and across international governments. ... This is a virus that attacked Windows platforms. ... The fact is the NHS has fallen victim to this. ... I don't think it's to do with that preparedness*" (The Independent, 2017). This message was likely meant to remove the gloom and potential public despair associated with a targeted NHS attack, as well as to reassure people that the country is prepared for such events. This aims to support people's trust in the government and in technological systems.

The psychological impact of WannaCry was also significant. For many it resulted in worry, anguish, disbelief, and a sense of helplessness – these are many of the issues discussed earlier in this chapter (Symantec, 2010; Böhme and Moore, 2012; Nurse, 2018). If we use the infection of NHS locations as an example, there are numerous cases to examine. Possibly one of the most well-reported in the set is that of a man who was preparing for a heart surgery and had it cancelled hours before it was due to take place (Fisher, Therrien, Hand & McCague, 2017). This led to his frustration, inconvenience to family members who travelled to stay close by, and bewilderment as to why anyone would wish to attack a hospital. Psychologically he was also impacted given that he had mentally prepared himself for the heart surgery being conducted on that date, and because it would now need to be rescheduled to some point in the future. These factors correlate with many of the emotional reactions to cybercrime discussed before (Symantec, 2010).

Psychologically, there was also the realisation by many that cyber-attacks could now cause the loss of life. As reported by the BBC, a member of NHS staff noted: "*Absolute carnage in the NHS today. Two Hyperacute stroke centres (the field I work in) in London have closed as of this afternoon. Patients will almost certainly suffer and die because of this*" (Fisher et al., 2017). This view is also reported by another person – an IT specialist – interviewed in the article: "*This kind of attack usually causes some inconvenience or financial loss [to] its victims but in this case it may well cause loss of life*". This is an important point as it relates to the severity of the threat (when considering PMT) but it can also cause further issues, given that individuals' behavior will not necessarily prevent this type of attack.

**4.2 Lloyds Banking Group Denial-of-Service (DoS) attack in 2017**
In January 2017, the Lloyds Banking Group and other banks in the UK fell victim to a denial-of-service (DoS) attack that persisted over a two-day period. DoS describes an attack where systems are bombarded with illegitimate data or requests, and therefore become unable to respond to legitimate requests (e.g., for a webpage or service access) in a timely manner (Nurse, 2018). In this instance, the financial institutions were attacked by a distributed DoS (DDoS) in which the illegitimate requests originated from multiple, dynamically changing locations. The Lloyds group were one of the most significantly targeted banks during this attack, and attack impacts included system unavailability and limited online service access for its customers.

At the broad social and societal levels, the attack affected millions of bank customers. According to reports, perpetrators attempted to block access to the bank's 20 million UK accounts (Collinson, 2017). The result was an impact on individuals and businesses, particularly their ability to login to online systems. As such, some customers would have been unable to view balances, make payments (e.g., for rent and bills), and conduct bank transfers (e.g., for necessary one-off transactions). Another pertinent factor here was that this type of DDoS attack targeting the banking sector had occurred in the past.

In 2015, a similar cyber-attack occurred which compromised the services of UK banks, RBS and Natwest (Collinson, 2015). A particularly worrying reality of that attack was the time in which it occurred, i.e., near to payday; this resulted in a maximum impact on some individuals and some widespread panic. One noteworthy difference and a key factor in the more recent attack was the number of days across which it took place. This extended period of two days (with some reports even suggesting it was longer for some customers) meant that not only was the reputation of the bank damaged, but also that customers' lives may have been seriously disrupted during that time.

To consider the response from government, one MP made the point that the cyber-attack was worrying for society and that more needed to be done. According to The Guardian, the then-MP stated: "*The attack on Lloyds was deeply troubling. Thousands of customers were affected by this, the latest in a long list of failures and breaches of banking IT systems. … As I have already pointed out, it is time to consider whether a single point of responsible for cyber risk in the financial services sector is now required*" (Collinson, 2017). This demonstrates some high-level concern for the finance sector broadly as a result of these types of cyber-attacks. There are also parallels for wider public confidence in technology. To consider the 2015 attack, one Natwest customer tweeted: "*Can't log in to #natwest yet again to check up on some transactions... and they want me to opt out of paper statements?! No chance*" (Collinson, 2015). This is poignant as it relates to how individuals perceive technology and risk, and how attacks can result in less trust in technology.

Analyzing the psychological impact of the Lloyds DDoS attack, it caused customers to be upset and frustrated – this was therefore mainly an emotional response. As reported by the BBC, one customer expressed: "*Haven't been able to access the site or app for over 36 hours now - is anything being done about this?*" (Peachey, 2017). This was one of a series of complains made on social media about the ongoing issue. Here we see one of the prime uses of social media (e.g., Twitter, Facebook and blogging platforms) today – i.e., allowing members of the public to directly reach out to companies (particularly for complains) and have their voice heard publicly.

While it is difficult to know the full extent of psychological impact, we can assume that lack of access to bank accounts and potentially personal funds (e.g., if money had to be transferred from one account to the another to facilitate a withdrawal), would have significantly increased customer stress and anxiety. There could have been several crucial reasons for account access, and depending on those needs, the emotions of stress, anger, pain, depression or helplessness could have resulted. This means that many of the follow-up issues highlighted in Section 2 and 3 immediately become relevant here. As mentioned earlier, the length of the disruption was a core factor as it would have exacerbated any initial inconvenience. These are all important concerns because they could each incite a change in how the public views cyber-attacks and whether they believe they have any control or skills to protect themselves or their families.

## 5. Conclusions

As online threats and cyber-attacks continue to permeate the Internet, it is essential that we as a community develop a better understanding of these issues and how they can impact our lives. This chapter took a significant step towards that goal by exploring how members of the public perceive and engage with risk and how they can be impacted after a cyber-attack has taken place. We focused on the social and psychological impacts of attacks as these are often overlooked in research and practice. These are, however, crucial factors in enhancing our understanding the broader side of attack impacts. To ground our work, we examined two well-known cyber-attacks and considered them in the context of the breadth of outcomes. It is expected that this research will motivate others to further investigate this area and the interaction between cybersecurity and cognitive factors.

**References**

Adams, J. (2013). Risk compensation in cities at risk. In: Joffe, H., Rossetto, T., Adams, J. (eds.) Cities at Risk. ANTHR, pp. 25–44. Springer, Netherlands.

Ajzen, I. (2002). Perceived Behavioral Control, Self-Efficacy, Locus of Control, and the Theory of Planned Behavior. Journal of Applied Social Psychology, 32, 665-683.

Altheide, D.L. (2002) Creating Fear; News and the Construction of Crisis, (Aldine De Gruyter; New York).: p.24.

Bada, M., Sasse, A.M. & Nurse, J. R. C. (2015) Cyber Security Awareness Campaigns: Why do they fail to change behaviour?, in proceedings of the International Conference on Cyber Security for Sustainable Society (CSSS) Coventry, UK, 118-131. SSN+.

Bandura, A. (1986). Fearful expectations and avoidant actions as coeffects of perceived self-inefficacy. American Psychologist, 41(12), 1389-1391.

Bandura, A. (1991). Social cognitive theory of self-regulation. Organizational Behavior and Human Decision Processes, 50, 248-287.

Bandura, A. & Adams N.E., (1977). Analysis of Self-Efficacy Theory of Behavioral Change. Cognitive Therapy and Research, Vol. 1, No. 4, pp. 287-310.

Barlyn, S. (2017). Global cyber attack could spur $53 billion in losses - Lloyd's of London. Reuters. Retrieved July 14 2018, from https://uk.reuters.com/article/uk-cyber-lloyds-report/global-cyber-attack-could-spur-53-billion-in-losses-lloyds-of-london-idUKKBN1A20AH

BBC. (2017a). NHS 'robust' after cyber-attack Retrieved July 14 2018, from https://www.bbc.co.uk/news/uk-39909441

Beck, U. (1999). World Risk Society. Cambridge: Polity Press.

Betz, D. J. & Stevens, T. (2011). Cyberspace and the State. London: Routledge.

Furedi, F. (2002). Culture of fear: Risk-taking and the morality of low expectation. London: Continuum.

Blythe, J., Camp, J. & Garg, V. (2011). Targeted risk communication for computer security, in 15th International Conference on Intelligent User Interfaces, pp. 295–298.

Blythe J. & Camp, J. L. (2012). Implementing mental models. IEEE Symposium on Security and Privacy Workshops, 24-25 May 2012, San Francisco, CA, 86-90.

Böhme, R. & Moore, T. (2012). How do consumers react to cybercrime? eCrime Researchers Summit, Las Croabas, pp. 1-12.

Collinson, P. (2015). Cyber attack hits RBS and NatWest online customers on payday. The Guardian. Retrieved July 4 2018, from https://www.theguardian.com/business/2015/jul/31/rbs-and-natwest-customers-complain-of-online-problemss

Collinson, P. (2017). Lloyds bank accounts targeted in huge cybercrime attack. The Guardian. Retrieved July 4 2018, from https://www.theguardian.com/business/2017/jan/23/lloyds-bank-accounts-targeted-cybercrime-attack

Dallaway, E. (2016). #ISC2Congress: Cybercrime Victims Left Depressed and Traumatized. Infosecurity Magazine. Retrieved July 4 2018, from https://www.infosecurity-magazine.com/news/isc2congress-cybercrime-victims/

Dickert, S., Västfjäll, D., Mauro, R., & Slovic, P. (2015). The feeling of risk: Implications for risk perception and communication. In H. Cho, T. Reimer, & K. A. McComas (Eds.), The SAGE handbook of risk communication (pp. 41–54). Thousand Oaks, CA: Sage Publications.

Fisher, M., Therrien, A., Hand, J. & McCague, B. (2017). How cyber-attack is disrupting NHS. BBC News. Retrieved July 4 2018, from https://www.bbc.com/news/live/39901370

Furedi, F. (2002). Culture of fear: Risk-taking and the morality of low expectation. London: Continuum.

Hale, C. (1996). Fear of crime: a review of the literature. International Review of Victimology 4:79–150.

Hirtz, Rob. (1998). Martin Seligman's journey from learned helplessness to learned happiness. The Pennsylvania Gazette. Retrieved August 4 2018, from http://www.upenn.edu/gazette/0199/hirtz.html

Gandhi, R., Sharma, A., Mahoney, W., Sousan, W., Zhu, Q. & Laplante, P. (2011). Dimensions of cyber attacks: Social, political, economic, and cultural. IEEE Technology & Society Magazine, 30(1), 28-38.

Garland, D. (2001) The Culture of Control; Crime and Social Order in Contemporary Society, OUP: Oxford.

Gross, M. L., Canetti, D., & Vashdi, D.R. (2016). The psychological effects of cyber terrorism. Bulletin of the Atomic Scientists, 72(5), 284-291.

Gross, M.L., Canetti, D., Vashdi, D.R. (2017) Cyberterrorism: its effects on psychological well-being, public confidence and political attitudes, Journal of Cybersecurity, 3(1), 49–58.

Iuga, C., Nurse, J. R. C., & Erola, A. (2016). Baiting the hook: factors impacting susceptibility to phishing attacks. Human-centric Computing and Information Sciences, 6(8). https://doi.org/10.1186/s13673-016-0065-2

Kirwan, G. & Power, A. (2011). The Psychology of Cyber Crime: Concepts and Principles. IGI Global.

Kleinot, M.C., & Rogers, R.W. (1982). Identifying effective components of alcohol misuse prevention programs. Journal of Studies on Alcohol, 43, 802-811.

Lawson, S. (2013). Beyond cyber-doom: Assessing the limits of hypothetical scenarios in the framing of cyber-threats. Journal of Information Technology & Politics 10(1), 86-103.

Leventhal, H. (1970). Findings and theory in the study of fear communications. In L. Berkowitz (Ed.), Advances in experimental social psychology. 5:119-186. New York: Academic Press.

Lynn, R. (2007). Virtual Rape is Traumatic, but is it a Crime? Retrieved August 4 2018, from http://www.wired.com/culture/lifestyle/commentary/sexdrive/2007/05/sexdrive_0504

Maddux, J.E., & Rogers, R.W. (1983). Protection motivation and self-efficacy: A revised theory of fear appeals and attitude change. Journal of Experimental Social Psychology, 19, 469-479.

Minei, E. & Matusitz, J. (2011). Cyberterrorist messages and their effects on targets: A qualitative analysis. Journal of Human Behaviour in the Social Environment, 21(8), 995-1019.

Modic, D., & Anderson, R. (2015). It's All Over but the Crying: The Emotional and Financial Impact of Internet Fraud. IEEE Security & Privacy, 13(5), 99-103.

Nurse, J. R. C. (2018). Cybercrime and You: How Criminals Attack and the Human Factors that They Seek to Exploit. In Attrill-Smith, A., Fullwood, C. Keep, M. & Kuss, D.J. (Eds.), Oxford Handbook of Cyberpsychology 2nd Edition. Oxford: OUP. https://doi.org/10.1093/oxfordhb/9780198812746.013.35

Nurse, J. R. C. & Bada, M. (2018). The Group Element of Cybercrime: Types, Dynamics, and Criminal Operations. In Attrill-Smith, A., Fullwood, C. Keep, M. & Kuss, D.J. (Eds.), Oxford Handbook of Cyberpsychology 2nd Edition. Oxford: OUP. https://doi.org/10.1093/oxfordhb/9780198812746.013.36

Nurse, J. R. C., Creese, S., & De Roure, D. (2017). Security risk assessment in Internet of Things systems. IT Professional, 19(5), 20-26. IEEE. https://doi.org/10.1109/MITP.2017.3680959

Nurse, J. R. C., Creese, S., Goldsmith, M., & Lamberts, K. (2011). Trustworthy and effective communication of cybersecurity risks: A review. In Proceedings of International Workshop on Socio-Technical Aspects in Security and Trust (STAST), pp. 60-68. IEEE. https://doi.org/10.1109/STAST.2011.6059257

Peachey, K. (2017). Lloyds online banking problems enter second day. BBC News. Retrieved August 14 2018, from https://www.bbc.co.uk/news/business-38594058

Prochaska, J.O., Redding, C.A., & Evers, K. (2002). The Transtheoretical Model and Stages of Change. In K. Glanz, B.K. Rimer & F.M. Lewis, (Eds.) Health Behavior and Health Education: Theory, Research, and Practice (3rd Ed.). San Francisco, CA: Jossey-Bass, Inc.

Reeve, T. (2017). Once bitten, twice shy: ONS stats reveal public response to cyber-crime. Magazine. Retrieved August 14 2018, from https://www.scmagazineuk.com/once-bitten-twice-shy-ons-stats-reveal-public-response-cyber-crime/article/1475468

Reid, L. W., Roberts, J. T., & Hilliard, H. M. (1998). Fear of crime and collective action: An analysis of coping strategies. Sociological Inquiry, 68(3), 312-328.

Reuters. (2017). Cyber attack hits 200,000 in at least 150 countries: Europol. Retrieved August 14 2018, from https://www.reuters.com/article/us-cyber-attack-europol/cyber-attack-hits-200000-in-at-least-150-countries-europol-idUSKCN18A0FX

Rippetoe, P.A., & Rogers, R.W. (1987). Effects of components of protection-motivation theory on adaptive and maladaptive coping with a health threat. Journal of Personality and Social Psychology, 52, 596-604.

Rogers, R.W. (1975). A protection motivation theory of fear appeals and attitude change. Journal of Psychology, 91, 93-114.

Rogers, R.W., & Mewborn, C.R. (1976). Fear appeals and attitude change: Effects of a threat's noxiousness, probability of occurrence, and the efficacy of the coping responses. Journal of Personality and Social Psychology, 34, 54-61.

Rogers, R.W., & Prentice-Dunn, S. (1997). Protection motivation theory. In D. Gochman (Ed.), Handbook of health behavior research: Vol. 1. Determinants of health behavior: Personal and social (pp. 113-132). New York, NY: Plenum.

Rogers, M. B., Amlôt, R., Rubin, G., Wessely, S. & Krieger, K. (2007). Mediating the Social and Psychological Impacts of Terrorist Attacks: The Role of Risk Perception and Risk Communication. International Review of Psychiatry, 19, 279-288.

Sanger, D.E., Chan, S. and Scott, M. (2017). Ransomware's Aftershocks Feared as U.S. Warns of Complexity. The New York Times. Retrieved July 14 2018, from https://www.nytimes.com/2017/05/14/world/europe/cyberattacks-hack-computers-monday.html

Seligman, Martin E. P. (1975). Helplessness: On Depression, Development, and Death. San Francisco: W.H. Freeman.

Slovic, P. (1988). Risk perception. In C. C. Travis (Ed.), Contemporary issues in risk analysis: Vol. 3: Carcinogen risk assessment (pp. 171-181). New York: Plenum.

Slovic, P. (2000). The perception of risk. London: Earthscan.

Sjöberg, L. (2000). Factors in Risk Perception. Risk Analysis, 20: 1-12.

Stekel, W. (1930). Les etats d' angoisse nerveux. Payot.

Suler, J. (2004). The Online Disinhibition Effect. Cyberpsychology and behaviour, 7(3), 321-324.

Sutherland, S. (2007). Irrationality: The Enemy Within. London: Pinter & Martin.

Symantec. (2010). Norton Cybercrime Report: The Human Impact. Retrieved June 14 2018, from https://www.symantec.com/content/en/us/home_homeoffice/media/pdf/cybercrime_report/Norton_USA-Human%20Impact-A4_Aug4-2.pdf

The Independent. (2017). NHS cyber attack: International manhunt to find criminals behind WannaCry ransomware that crippled hospital systems. Retrieved July 14 2018, from https://www.independent.co.uk/news/uk/home-news/wannacry-wanna-detector-accident-and-emergency-patient-appointment-operation-a7734831.html

Williams, E. J., Hinds, J., & Joinson, A. N. (2018). Exploring susceptibility to phishing in the workplace. International Journal of Human-Computer Studies, 120, 1-13.

Williams, M., Nurse, J. R. C., & Creese, S. (2017). Privacy is the Boring Bit: User Perceptions and Behaviour in the Internet-of-Things. In Proceedings of the 15th International Conference on Privacy, Security and Trust (PST). https://doi.org/10.1109/PST.2017.00029

Williams, M., Axon, L., Nurse, J. R. C., & Creese, S. (2016). Future scenarios and challenges for security and privacy. In Research and Technologies for Society and Industry Leveraging a better tomorrow (RTSI), 2016 IEEE 2nd International Forum on (pp. 1-6). IEEE. https://doi.org/10.1109/RTSI.2016.7740625

Witte, K. (1991). Preventing AIDS through persuasive communication: Fear appeals and preventive-action efficacy. Doctoral dissertation, University of California, Irvine.

Witte, K. (1992). The role of threat and efficacy in AIDS prevention. International Quarterly of Community Health Education, 12, 225-249.

Witte, K. (1992a). Putting the Fear Back in Fear Appeals: The Extended Parallel Process Model. Communication Monographs, 59, 329-349.

Witte, K. (1994). Fear control and danger control: A test of the Extended Parallel Process Model (EPPM). Communication Monographs, 61, 113-134.

Verizon. (2018). 2018 Data Breach Investigations Report. Retrieved August 4 2018, from https://www.verizonenterprise.com/verizon-insights-lab/dbir/

Virtanen, S. (2017). Fear of Cybercrime in Europe: Examining the Effects of Victimization and Vulnerabilities, Psychiatry, Psychology and Law, 24:3, 323-338.